# Outstandingly high thermal conductivity, elastic modulus, carrier mobility and piezoelectricity in two-dimensional semiconducting CrC$_2$N$_4$: A first-principles study

Bohayra Mortazavi*,a,b, Fazel Shojaeic,#, Brahmanandam Javvajia,#, Timon Rabczukd and Xiaoying Zhuang**a, d

aChair of Computational Science and Simulation Technology, Institute of Photonics, Department of Mathematics and Physics, Leibniz Universität Hannover, Appelstraße 11,30167 Hannover, Germany.
bCluster of Excellence PhoenixD (Photonics, Optics, and Engineering–Innovation Across Disciplines), Gottfried Wilhelm Leibniz Universität Hannover, Hannover, Germany.
cDepartment of Chemistry, Faculty of Sciences, Persian Gulf University, Boushehr 75168, Iran.
dCollege of Civil Engineering, Department of Geotechnical Engineering, Tongji University, 1239 Siping Road Shanghai, China.

**Abstract**

Experimental realization of single-layer MoSi$_2$N$_4$ is among the latest groundbreaking advances in the field of two-dimensional (2D) materials. Inspired by this accomplishment, herein we conduct first-principles calculations to explore the stability of MC$_2$N$_4$ (M= Cr, Mo, W, V, Nb, Ta, Ti, Zr, Hf) monolayers. Acquired results confirm the desirable thermal, dynamical and mechanical stability of MC$_2$N$_4$ (M= Cr, Mo, W, V) nanosheets. Interestingly, CrC$_2$N$_4$, MoC$_2$N$_4$ and WC$_2$N$_4$ monolayers are found to be semiconductors with band gaps of 2.32, 2.76 and 2.86 eV, respectively, using the HSE06 functional, whereas VC$_2$N$_4$ lattice shows a metallic nature. The direct gap semiconducting nature of CrC$_2$N$_4$ monolayer results in excellent absorption of visible light. The elastic modulus and tensile strength of CrC$_2$N$_4$ nanosheet are predicted to be remarkably high, 676 and 54.8 GPa, respectively. On the basis of iterative solutions of the Boltzmann transport equation, the room temperature lattice thermal conductivity of CrC$_2$N$_4$ monolayer is predicted to be 350 W/mK, among the highest in 2D semiconductors. CrC$_2$N$_4$ and WC$_2$N$_4$ lattices are also found to exhibit outstandingly high piezoelectric coefficients. This study introduces CrC$_2$N$_4$ nanosheet as a novel 2D semiconductor with outstandingly high mechanical strength, thermal conductivity, carrier mobility and piezoelectric coefficient.

Keywords: *2D materials; piezoelectric; mechanical; thermal conductivity; semiconductor*

## 1. Introduction

In line of continuous endeavors in the fabrication of novel two-dimensional (2D) semiconductors, Hong and coworkers [1] most recently devised an innovative strategy and could fabricate large-area $MoSi_2N_4$ monolayer, by including silicon during the growth of molybdenum nitride via the chemical vapor deposition technique. This experimental achievement is believed to be among the latest greatest groundbreaking advances in the field of 2D materials [2]. First of all, the proposed experimental approach can be employed to fabricate an entirely new class of 2D materials with $MA_2Z_4$ formula, where M can be an early transition metal (Mo, W, V, Nb, Ta, Ti, Zr, Hf or Cr), A atom can take Si or Ge and Z atom is N, P or As [2]. Theoretical studies confirm that $CrSi_2N_4$, $MoSi_2N_4$ and $WSi_2N_4$ nanosheets are semiconductors with high carrier mobilities [1,3] and can exhibit remarkably high mechanical and thermal conduction properties and exceptional piezoelectricity [3]. One challenging aspect that arises here is that, if carbon atoms get incorporated instead of silicon during the chemical growth of an early transition metal nitride, is the resulting $MC_2N_4$ nanosheet stable and if yes, what are the expectable properties.

Density functional theory calculations provide an efficient possibility to design and predict novel materials with appealing physical properties [4–6]. Motivated by the experimental work by Hong *et. al*, [1] in this work our objective is to explore the stability and intrinsic properties of $MC_2N_4$ (M= Cr, Mo, W, V, Nb, Ta, Ti, Zr, Hf) monolayers, on the basis of first-principles density functional theory calculations. We first investigate the structural and bonding mechanism in $MC_2N_4$ nanomembranes. Next the dynamical and thermal stability of considered monolayers are systematically examined by calculating the phonon dispersion relations and ab-initio molecular dynamics simulations, respectively. For the thermally and dynamically stable lattices we then elaborately explore their electronic, mechanical and heat transport properties. In depth analysis of phonon transport and lattice thermal conductivity are accomplished on the basis of iterative solution of the Boltzmann transport equation. The research concerning the piezoelectric properties of nanomaterials and particularly 2D materials is highly increasing and is beneficial to harvest energy for ultra-small devices [7–9]. During the mechanical deformation of 2D materials, bond stretching and bond bending can change the local electric effects that may lead to formation of dipole moment or polarization. The introduction of asymmetry to nanosheets by changing the type atoms in one side (Janus structures) can create an intrinsic dipole moment and substantially enhance piezoelectric coefficient [9–14]. For the semiconducting lattices, we thus also calculate the piezoelectric coefficients. Our extensive results reveal outstandingly high elastic modulus, lattice thermal conductivity, carrier mobility and piezoelectric coefficient of thermally and dynamically stable $CrC_2N_4$ nanosheet, highly motivating for the application in nanoelectronics, thermal management and energy storage/conversion systems.

## 2. Computational methods

We conduct density functional theory (DFT) calculations with the generalized gradient approximation (GGA) and Perdew–Burke–Ernzerhof (PBE) [15], as implemented in *Vienna Ab-initio Simulation Package* [16,17]. Projector augmented wave method was used to treat the electron-ion interactions [18,19] with a cutoff energy of 500 eV for the plane waves with energy convergence criteria of $10^{-5}$ eV. For the geometry optimizations, atoms and lattices are relaxed according to the Hellman-Feynman forces using conjugate gradient algorithm until atomic forces drop to lower than 0.001 eV/Å [20]. The first Brillouin zone (BZ) was sampled with 14×14×1 Monkhorst-Pack [21] k-point grid. Stress-strain relations are examined by conducting uniaxial tensile loading simulations. Since PBE/GGA functional systematically underestimates the electronic band gaps, HSE06 hybrid functional [22] is employed to more precisely examine the electronic nature. Carrier mobilities are predicted on the basis of the deformation potential approximation [23] using $\frac{e\hbar^3 C_{2D}}{KT m_e^* m_d (E_l^i)^2}$, where K is the Boltzmann constant, $\hbar$ is the reduced Planck constant, $m_d$ is the average effective mass, $C_{2D}$ and $m^*$ are, respectively, the elastic modulus and the effective mass and $E_l^i$ is the carrier deformation energy constant for the i-th edge band phonons. Light absorption is reported on the basis of frequency-dependent dielectric matrix, without taking into account the local field effects. The imaginary part ($\varepsilon_2$) of dielectric matrix is obtained from the following equation:

$$\varepsilon_{\alpha\beta}^2(\omega) = \frac{4\pi^2 e^2}{\Omega} \lim_{q \to 0} \frac{1}{q^2} \sum_{c,v,\mathbf{k}} 2w_k \delta(\varepsilon_{c\mathbf{k}} - \varepsilon_{v\mathbf{k}} - \omega) <u_{c\mathbf{k}+\mathbf{e}_\alpha q} | u_{v\mathbf{k}}><u_{c\mathbf{k}+\mathbf{e}_\beta q} | u_{v\mathbf{k}}>^*$$

(1)

where indices *c* and *v* refer to conduction and valence band states, respectively; $w_k$ is the weight of the *k*-point; and $u_{ck}$ is the cell periodic part of the orbitals at the *k*-point. The real part ($\varepsilon_1$) of the tensor is obtained from the Kramers-Kronig relation [24]. The absorption coefficient is calculated from the following:

$$\alpha(\omega) = \sqrt{2}\omega \left[\frac{\sqrt{\varepsilon_1^2+\varepsilon_2^2}-\varepsilon_1}{2}\right]^{1/2} \quad (2)$$

We conduct density functional perturbation theory (DFPT) calculations for 5×5×1 supercells to acquire interatomic 2nd order harmonic force constants. Phonon dispersion relations and harmonic force constants are calculated using the PHONOPY code [25]. Moment tensor potentials (MTPs)[26] are trained to evaluate the phononic properties [27] using the MLIP package [28]. Ab-initio molecular dynamics (AIMD) simulations are conducted for 5×5×1 supercells with a time step of 1 fs [27]. The training sets for the development of MTPs are prepared by conducting two separate AIMD simulations from 20 to 100 K and 200 to 1000 K for 1000 and 1500 time steps, respectively. From every separate calculation, 500 trajectories were subsamples and used in the final training set. AIMD simulations are conducted at 500 and 1000 K for 20000 time steps over supercells with 80 atoms to examine the thermal stability. Anharmonic 3rd order interatomic force constants are obtained for 5×5×1 supercells using the trained MTPs by considering the interactions with eleventh nearest neighbors. Full iterative solution of the Boltzmann transport equation is carried out to estimate the phononic thermal conductivity using the ShengBTE [29] package with 2nd and 3rd order force constants inputs, as discussed in our earlier study [30]. In all calculations the isotope scattering is considered and the convergence of the thermal conductivity to the $q$-grid is examined and minimum of 41×41×1 $q$-grids are used.

The electrical polarization under the given mechanical loading is estimated using a combination of short-range MTP potential and the long-range charge-dipole (CD) interactions. The CD model [31,32] assumes that each atom is associated with a charge $q$ and **p** dipole moment. From the atomic polarizability $R$ and electron affinity $\chi$ of each atom type, we estimate $q$ and **p** by minimizing the CD potentials. $\chi$ values for different atom types are known from literature [33–35]. The unknown quantity $R$ obtained from the DFT estimated isotropic polarizability ($\alpha_{DFT}$) of small supercells. The total polarizability from the CD model ($\alpha_{CDM}$) is also evaluated for these supercells assuming that $R$ varies between 0.1 to 2.0 Å. Then establishing a close match between $\alpha_{DFT}$ and $\alpha_{CDM}$ results the parameter $R$. The GAUSSIAN software [36] is employed to estimate the $\alpha_{DFT}$. LAMMPS package [37] is used to perform the mechanical deformation simulations along with the CD model interactions. The detailed procedure of obtaining the CD parameters and linking to LAMMPS are discussed in earlier works [38,39].

Herein we consider a square sheet of $MC_2N_4$ monolayers of dimensions nearly 80x80 Å$^2$. The atomic configuration first thermally equilibrated to 0.1K using NVT time integration for 10 ps with a time step of 1 fs. Then atoms are displaced using the displacement field $u_y = K_t y$, where $K_t$ is the strain and $y$ represents the atomic coordinate in the $y$ direction. We hold the displacement of sheet edges and relax the atomic system for 1000 time-steps. During this period, the atoms redistribute the forces and stabilize the energy fluctuations. We then apply the displacement field again and relax the structure for another 1000 steps. This process continues until the atomic configuration reaches a strain of 0.015. During deformation, the numerical values of charge and dipole moments for each atom are estimated from the CD model equations. The total polarization of the atomic system is $\mathbf{P} = \frac{\sum_{i=1}^{n} \mathbf{p}_i}{A}$, where $A$ is the area of the monolayer and $n$ is the total number of moving atoms. The piezo coefficient is then extracted by estimating the linear relationship between polarization and strain fields.

## 3. Results and discussions

We first study the structural and stability of $MC_2N_4$ monolayers with two different configurations of 1T and 2H, in analogy to transition metal dichalcogenides lattices. Fig. 1a displays the crystal structure of a $MC_2N_4$ monolayer in two distinct phases of 1T and 2H. The monolayer in both phases has a hexagonal primitive cell with a consistent chemical formula per unitcell. As it can be seen in the figure, each $MC_2N_4$ monolayer comprises a 2H/1T-$MN_2$ layer covalently sandwiched in-between two eclipsed oriented CN layers. The difference between the two phases is solely related with different coordination geometry of M atom. Such that in 2H phase, M atom has a trigonal prismatic coordination to six N atoms while in 1T counterpart, M atom has an octahedral coordination. In this work, for central M atoms we consider early transition metals of Cr, Mo, W, V, Nb, Ta, Ti, Zr, and Hf to build $MC_2N_4$ monolayers.

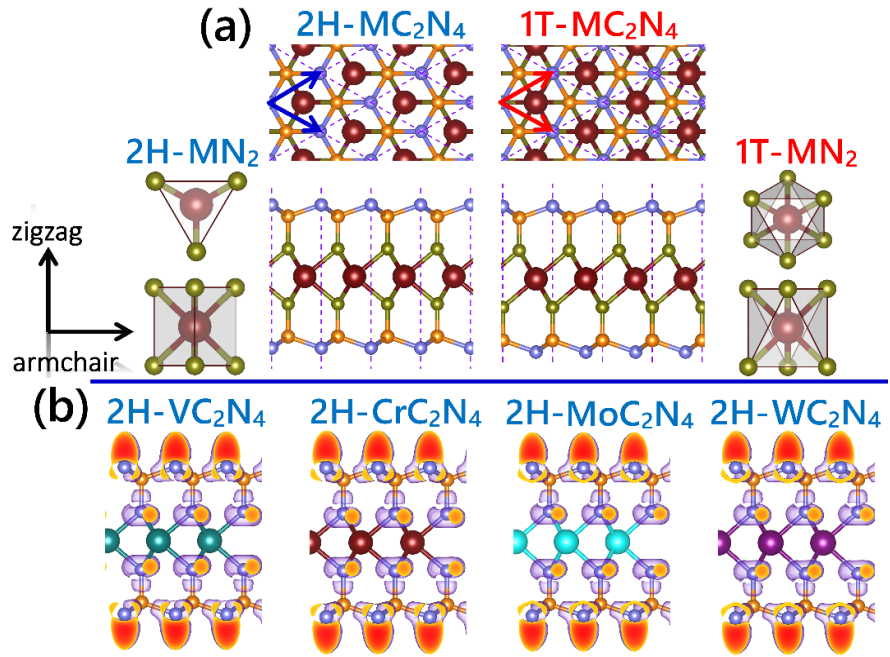

**Fig. 1**, (a) Top and side views of MC$_2$N$_4$ monolayer in 2H and 1T phases. The hexagonal primitive cell, lattice vectors, and the trigonal-prismatic coordination and octahedral coordination of M atom of MN$_2$ motif in 2H and 1T phases are also shown. (b) Electron localization function (ELF) calculated for 2H-MC$_2$N$_4$ (M= V, Cr, Mo, W) monolayers. In this figure light blue and green colors represent N atoms, wine color M atoms, and orange color C atoms.

From the energetic point of view, in accordance with MoSi$_2$N$_4$ family [3], we found that 2H structures exhibit lower energies than their 1T counterparts, except for the cases of TiC$_2$N$_4$ and ZrC$_2$N$_4$ monolayers. Nonetheless, in agreement with our earlier study finding for the case of MoSi$_2$N$_4$ family [3], we realized that 1T phases are dynamically unstable. We therefore focus only on the 2H lattices and further examine their stability. Worthy to note that our spin polarized calculations confirm that none of the considered 2H monolayers show magnetic moments. In Fig. 2 the predicted phonon dispersion relations of 2H-MC$_2$N$_4$ monolayers are presented. These results confirm that 2H-MC$_2$N$_4$ (M=Ti, V, Nb, Ta, Cr, Mo, W) are dynamically stable, as they do not exhibit imaginary frequencies. It is clear that only ZrC$_2$N$_4$ and HfC$_2$N$_4$ are dynamically unstable, as in the both cases acoustic modes show considerable imaginary frequencies. For the practical applications, the analysis of thermal stability is ought to be also conducted. AIMD results reveal that after 18 ps long simulations VC$_2$N$_4$ and CrC$_2$N$_4$ monolayers stay fully intact at 500 and 1000 K temperatures, confirming their excellent thermal stability (find Fig. S1). NbC$_2$N$_4$ and TaC$_2$N$_4$ monolayers nonetheless disintegrate at both 500 and 1000 K temperatures, which show their undesirable thermal stability. MoC$_2$N$_4$ and WC$_2$N$_4$ nanosheets are found to be highly stable at 500 K but distorted at 1000 K, thus they also show acceptable thermal stability (find Fig. S1). These results demonstrate that generally by increasing the weight of transition metals the thermal stability decreases. Moreover, the stability of these systems are directly related to their lattice constant, and such that $CrC_2N_4$ shows the maximum stability since it shows the smallest lattice constant and minimum C-N bond lengths in the top layer. By increasing the lattice constant, the C-N bonds stretching may result in either imaginary frequency in phonon dispersions or lattice distortions at high temperatures. From our originally considered 18 $MC_2N_4$ monolayers, it is clear that only 2H-$MC_2N_4$ (M= V, Cr, Mo, W) nanosheets might be experimentally producible and stable for practical applications. The geometry optimized lattices are included in the supplementary information document and structural and electronic properties are summarized in Table 1.

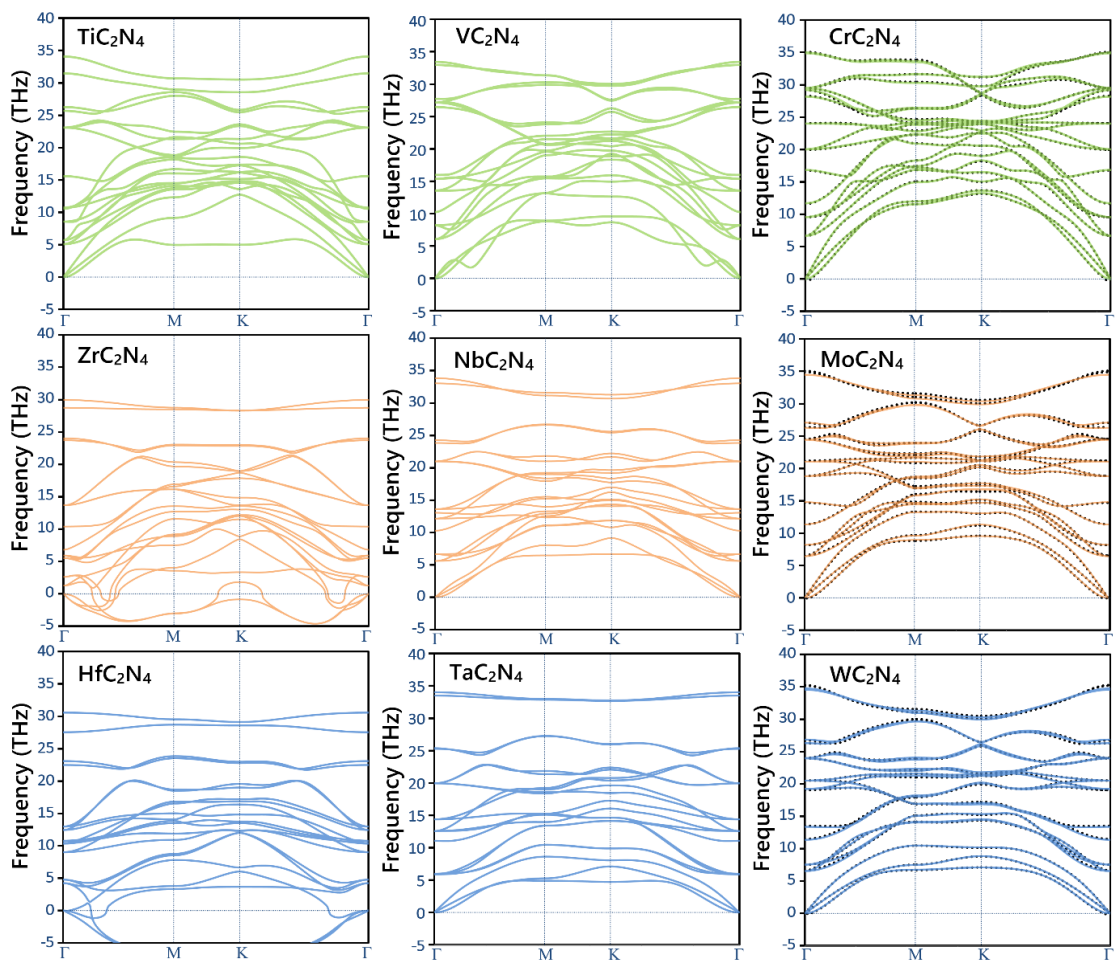

**Fig. 2**, Phonon dispersion relations of 2H-$MC_2N_4$ monolayer predicted using the MTP method. For the cases of $MC_2N_4$ (M= Cr, Mo, W) monolayers the DFPT results are also shown by the dotted lines.

We next study the bonding mechanism in dynamically and thermally stable 2H-$MC_2N_4$ (M= V, Cr, Mo, W) monolayers. These nanomembranes exhibit a rather complex crystal structure with different types of chemical bonds along the in-plane and out-of-plane directions. We perform Bader charge analysis [40] as well as electron localization function (ELF) [41] to investigate the nature of chemical bonds in these monolayers. As summarized in Table 1, in all cases M atom is positively charged (Cr: +1.25 e, V: +1.32 e, Mo: +1.34 e, W:+1.51), while each of surrounding N atoms, belonging to the MN$_2$ layer, is on average by 1.10 e negatively charged. This indicates that electrostatic interactions are dominating within MN$_2$ layer. The carbon atoms which are quadruply bonded to N atoms are also positively charged (on average +1.09 e) due to the higher electronegativity of N atoms. The calculated C-N distances in CN layer and that of those connecting MN$_2$ layer and CN layers are 1.57 (on average) and 1.42 Å, respectively, slightly different than that of a single covalent C-N bond (1.42 Å). The ELF results with isosurface value of 0.75 are depicted in Fig. 1(b). From Fig. 1(b), one can clearly see that electrons are strongly localized in between of each directly bonded C and N pairs indicating polar covalent C-N bonds, and also on N atoms of CN layers, which correspond to the lone pair electrons. In agreement with Bader analysis, the small ELF values in regions in between M and N atoms of MN$_2$ layer clearly indicates the absence of strong covalent interactions.

Table 1, Calculated structural and electronic properties of MC$_2$N$_4$ (M= V, Cr, Mo, W) monolayers.

| System | Lc (Å)[a] | Thk. (Å)[b] | ($q_M$,$q_N$)[c] | ($q_C$,$q_N$)[d] | Electronic structure [e] | Transition k-points[f] |
|---|---|---|---|---|---|---|
| VC$_2$N$_4$ | 2.552 | 9.497 | 1.32,-1.06 | 1.06,-0.66 | $E_g^{PBE}$=0 eV (M) | - |
| CrC$_2$N$_4$ | 2.513 | 9.424 | 1.25,-1.06 | 1.16,-0.72 | $E_g^{PBE}$=1.78 eV (QD), $E_g^{HSE}$=2.32 eV (D) | (K →K) |
| MoC$_2$N$_4$ | 2.622 | 9.709 | 1.34,-1.11 | 1.09,-0.64 | $E_g^{PBE}$=1.80 eV (I), $E_g^{HSE}$=2.76 eV (I) | (K →Γ) |
| WC$_2$P$_4$ | 2.639 | 9.710 | 1.51,-1.20 | 1.06,-0.62 | $E_g^{PBE}$=1.86 eV (I), $E_g^{HSE}$=2.86 eV (I) | (K → Γ) |

[a]PBE/GGA optimized lattice constants. MC$_2$N$_4$ monolayers exhibit a hexagonal crystal lattice (a=b). [b]Effective thickness of a MC$_2$N$_4$ monolayer is defined as the sum of the monolayer apparent thickness plus Van der Waals diameter of N atoms. [b]Bader charge on metal atom ($q_M$) and its coordinated N atoms ($q_N$). [d]Bader charge on C ($q_C$) and N ($q_N$) atoms of CN layers. [e]PBE/GGA and HSE06 calculated band gaps. Abbreviations "I" and "D" indicate indirect and direct gap semiconducting character, respectively, and "M" indicate metallic character. [f]band gap transition k-points of HSE06 results.

We next shift our attention to the electronic properties of 2H-MC$_2$N$_4$ (M= V, Cr, Mo, W) monolayers. Fig. 3 depicts PBE and HSE06 predicted electronic band structures and projected density of states (PDOS) of aforementioned monolayers. From the presented results, it is apparent that the electronic structure of MC$_2$N$_4$ monolayer sensitively varies with type of M atom. It is found that CrC$_2$N$_4$ monolayer is a direct gap semiconductor with HSE06 (PBE) band gap of 2.32 (1.78) eV at K-point. As expected the general features of PBE and HSE06 band structures of CrC$_2$N$_4$ monolayer are consistent. Nonetheless, although not easily distinctive in Fig. 3, but the valance band maximum (VBM) of PBE band structure is slightly off-K-point, along the K-M. The calculated band gap is slightly larger than that obtained for 2H-MoSi$_2$N$_4$ (2.23 eV) [3], the experimentally realized prototype of MA$_2$Z$_4$ compounds. Both MoC$_2$N$_4$ and WC$_2$N$_4$ monolayers, however, exhibit indirect-gap nature with HSE06 (PBE) band gaps of 2.76 (1.80) and 2.86 (1.86) eV, respectively. For both compounds, HSE06 VBM is at K-point, while the conduction band minimum (CBM) occurs along the K-Γ direction. Concerning VC$_2$N$_4$ monolayer, one can simply deduce even without calculations that nonmagnetic VC$_2$N$_4$ must exhibit metallic behavior because of the odd number of electrons in its primitive cell. To better understand the nature of band edge states and also to rationalize the observed differences, we calculated atom-type projected band structures and PDOS (Fig. 3), orbital PDOS (shown in Fig. S2), as well as charge density distributions at VB(K), CB(K), and CB(K-Γ) (depicted in Fig. S3) for MC$_2$N$_4$ monolayers. For each system, VBM is almost dominantly made of M in-plane ($d_{x^2-y^2}, d_{xy}$) orbitals with a minor contribution from N($p_x, p_y, p_z$) orbitals, representing a bonding σ(M-M) state. As a result of strong interaction between d orbitals of neighboring M atoms, in all cases valance band is highly dispersed which leads to small hole effective masses and high hole mobility. The lattice constants values of MoC$_2$N$_4$ and WC$_2$N$_4$ monolayers are close and slightly different than that of the CrC$_2$N$_4$ monolayer. Therefore, M-M interaction is much more pronounced in them than CrC$_2$N$_4$, as evidenced by their steeper and slightly deeper (-7.71 and -7.68 eV for MoC$_2$N$_4$ and WC$_2$N$_4$ vs -7.30 eV for CrC$_2$N$_4$) valence bands (Fig. 3). The CBM in CrC$_2$N$_4$ monolayer, however, is solely made of M($d_{z^2}$) orbitals. Apparently, $d_{z^2}$ orbitals of neighboring M atoms make lesser interactions compared to the in-plane ($d_{x^2-y^2}, d_{xy}$) orbitals, leading to an almost flat conduction band(Fig. 2). For both MoC$_2$N$_4$ and WC$_2$N$_4$ monolayers, CBM is derived from M in-plane ($d_{x^2-y^2}, d_{xy}$) orbitals hybridized with N(s,pz) orbitals. Considering the fact that d orbitals of Mo and W are higher in energy than those of Cr atom and also VBM of MoC$_2$N$_4$ and WC$_2$N$_4$ are deeper in comparison with CrC$_2$N$_4$ monolayer, the larger band gaps of MoC$_2$N$_4$ and WC$_2$N$_4$ compounds can be clearly explained.

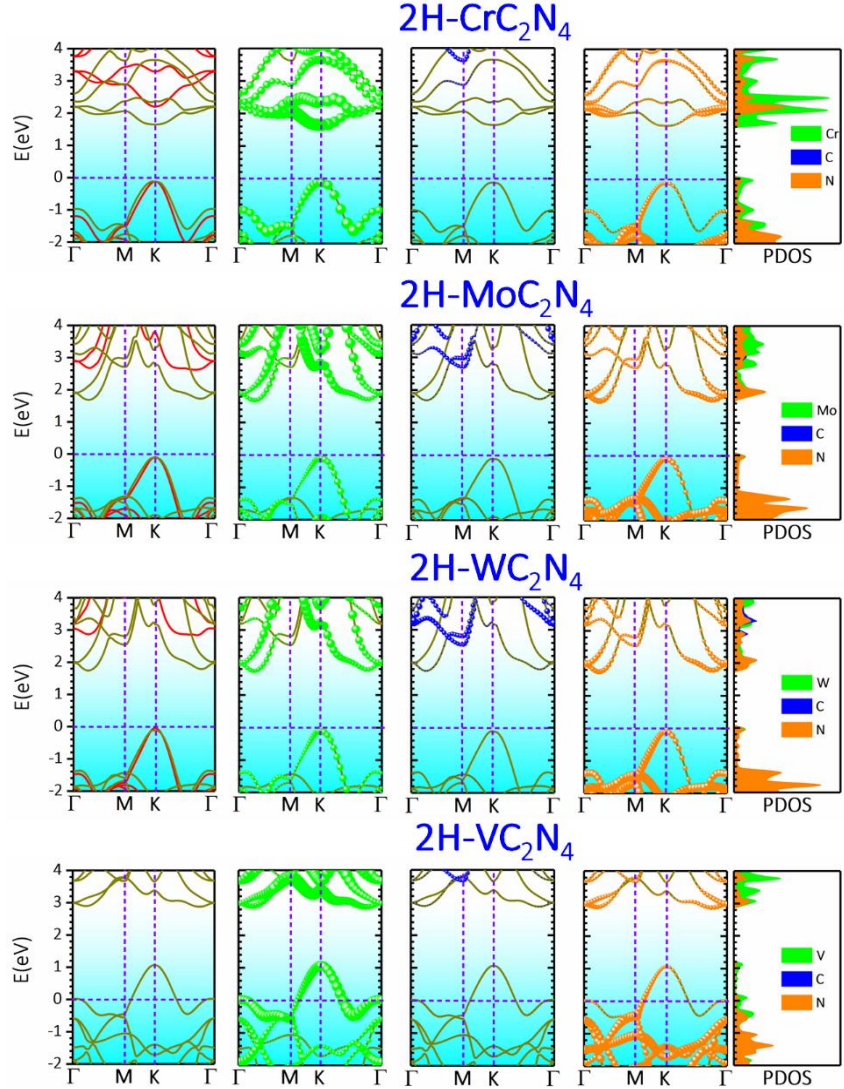

Fig. 3, (a) PBE and HSE06 band structures, atom-type projected band structures, and projected density of states (PDOS) calculated for 2H-$MC_2N_4$ (M= V, Cr, Mo, W) monolayers. In this figure red and green lines represent HSE06 and PBE based predicted electronic band structures. Green, blue, and brown filled dots also represent contribution of M, C, and N atoms to each band.

We then investigate the mechanical properties of thermally and dynamically stable $MC_2N_4$ (M= V, Cr, Mo, W) monolayers by evaluating the uniaxial stress-strain relations. In these calculations the stresses along the two perpendicular directions of the loading are ensured to stay negligible during various stages of the loading. The predicted uniaxial stress-strain relations of $MC_2N_4$ (M= V, Cr, Mo, W) monolayers along armchair and zigzag direction (as distinguished in Fig. 1) are illustrated in Fig. 4. In these results, the stresses values at every strain are calculated by considering the actual volume of the monolayers. To this goal the actual thickness at every strain level is measured as the normal distance between two boundary N atoms plus van der Waals diameter of N atoms (equal to 3.1 Å). Likely to the graphene and other densely packed 2D materials, the uniaxial stress-strain curves start with a linear relation associated with the linear elasticity, followed by a nonlinear trend up to the maximum tensile strength point. The elastic modulus is predicted by fitting a line to the stress-strain values up to the strain level of 0.01. Likely to the $MoSi_2N_4$ family [3], $MC_2N_4$ (M= V, Cr, Mo, W) nanosheets also exhibit isotropic elasticity, meaning that the elastic modulus along the armchair and zigzag directions are equal. The elastic modulus of $CrC_2N_4$, $MoC_2N_4$, $WC_2N_4$ and $VC_2N_4$ monolayers are estimated to be 676, 543, 553 and 577 GPa, respectively. It is conspicuous that these nanosheets show distinctly higher elastic modulus than their $MSi_2N_4$ counterparts [3]. Owing to strong covalent N-C bonds, $CrC_2N_4$ nanosheet shows around 44% higher elastic modulus than $CrSi_2N_4$ counterpart. In sharp contrast with elastic response, studied nanosheets exhibit highly anisotropic mechanical response and along zigzag they show remarkably higher tensile strength and stretchability than the armchair direction. The maximum tensile strength of $CrC_2N_4$, $MoC_2N_4$, $WC_2N_4$ and $VC_2N_4$ along the zigzag(armchair) directions are found to be 54.8(34.2), 37.0(18.3), 37.4(18.0) and 44.5(24.5) GPa, respectively. In accordance with structural and electronic properties, it is noticeable that $MoC_2N_4$ and $WC_2N_4$ nanomembranes show remarkably close mechanical properties and noticeably lower than their $CrC_2N_4$ and $VC_2N_4$ counterparts. As discussed earlier, $CrC_2N_4$ monolayer exhibits the smallest lattice constant and less stretched C-N bonds, which explains its superior mechanical properties than other considered structures.

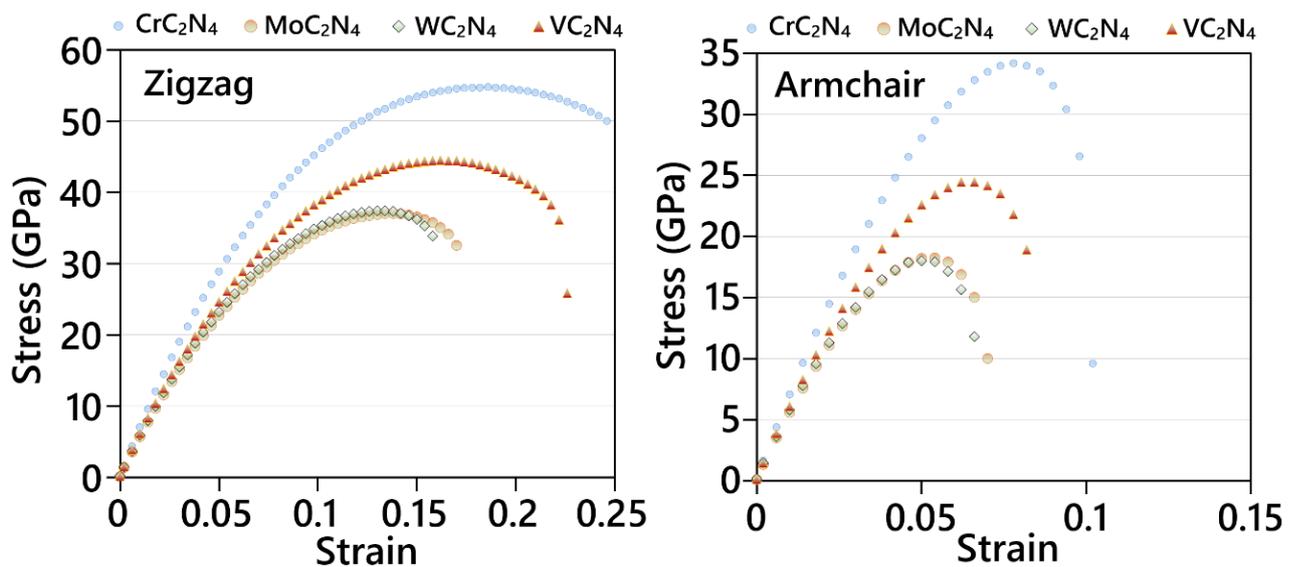

**Fig. 4**, Uniaxial stress-strain curves of $MC_2N_4$ monolayers along the zigzag and armchair and directions.

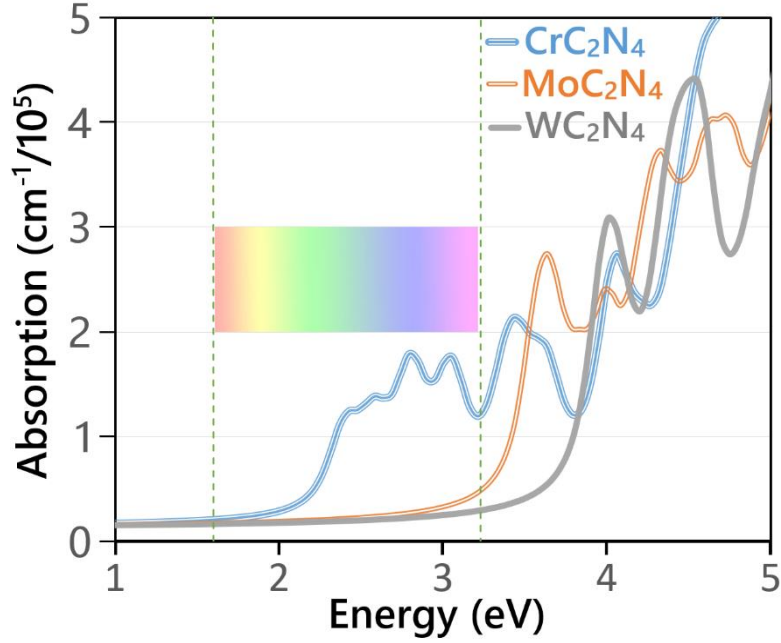

Fig. 5, Absorption spectra of $MC_2N_4$ (M= Cr, Mo, W) monolayers calculated using HSE06 functional. The visible-light energy range is also shown by vertical lines.

We next explore charge carrier mobility for semiconducting $MC_2N_4$ monolayers using the deformation potential theory (DPT) by assuming the absence of defects or any other external sources of scattering. For the convenience, the smallest rectangular lattice of each monolayer is used to calculate the electron and hole mobilities along the armchair and zigzag directions. Table 2 summarizes the estimated mobilities for $MC_2N_4$ (M= Cr, Mo, W) monolayers. It is conspicuous that in these systems, electron effective masses exhibit severe anisotropic behavior, while the hole effective masses as well as VBM and CBM deformation potentials weakly depend on direction. This leads to having anisotropic electron mobilities and weakly anisotropic hole mobilities along zigzag and armchair directions in these structures. Table 2 also reveals that the hole mobility is dramatically larger than the electron mobility along both zigzag and armchair directions in these monolayers, indicating that they exhibit p-type character. For the case of single-layer $CrC_2N_4$, the highest hole mobility is found to be 1235 $cm^2V^{-1}s^{-1}$, which is by around two order larger than its electron mobility (13 $cm^2V^{-1}s^{-1}$). A similar mobility trend was also observed for 2H-$MoSi_2N_4$ in our previous work [3], for which the highest carrier mobility (1100 $cm^2V^{-1}s^{-1}$) is found to be slightly smaller than that of $CrC_2N_4$. The highest hole mobilities in $MoC_2N_4$ and $WC_2N_4$ are found to be 8508 and 8424 $cm^2V^{-1}s^{-1}$, respectively, even larger than that in $CrC_2N_4$ by about 7 times. The hole mobility values are however appreciably smaller than those calculated for other N-rich 2D materials like $BeN_2$ monolayer [5]. The huge difference between electron and hole mobilities in these systems may enhance the separation of photogenerated electron-hole pairs which can enhance efficiency of optoelectronic devices.

To investigate possible application of 2H-MC$_2$N$_4$ (M= Cr, Mo, W) monolayers in optoelectronic devices, we used HSE06 functional to calculate light absorption coefficients ($\alpha(\omega)$), shown in Fig. 5. These monolayers are found to exhibit isotropic absorption coefficients in response to the light polarized along zigzag and armchair directions. The first absorption peak of CrC$_2$N$_4$ appears in the visible range at around 2.43 eV, slightly higher than its HSE06 band gap. It can be seen that CrC$_2$N$_4$ show remarkably high absorption coefficients ($10^5$ cm$^{-1}$) in a broad range of visible light frequencies, which are comparable to those of halide perovskites [42]. The first absorption peaks of MoC$_2$N$_4$ and WC$_2$N$_4$, however, appear at 3.85 and 4.27 eV, respectively, indicating these two compounds only show absorption in ultraviolet and higher frequencies. These first peak values are appreciably larger than HSE06 calculated band gaps for these two compounds, which can be attributed to their indirect nature of band gap. With a direct band gap of 2.23 eV, a maximum carrier mobility of 1235 cm$^2$V$^{-1}$s$^{-1}$, and large absorption coefficients in visible range, CrC$_2$N$_4$ can be a promising candidate for nanoelectronics and optoelectronic applications.

**Table 2**. Elastic modulus (C$_{2D}$), effective mass of electrons and holes ($m_e^*$, $m_h^*$) with respect to the free-electron mass (m$_0$), deformation energy of the CBM and VBM ($E_l^{CBM}$ and $E_l^{VBM}$), and mobility of electrons and holes ($\mu_e$, $\mu_h$) along zigzag and armchair directions for calculated for MC$_2$N$_4$ (M= V, Cr, Mo, W) monolayers.

|  |  |  | Electron | | | Hole | | |
|---|---|---|---|---|---|---|---|---|
| System | Direction | C$_{2D}$(N/m) | $m_e^*$/m$_0$ | $E_l^{CBM}$(eV) | $\mu^a$ | $m_h^*$/m$_0$ | $E_l^{VBM}$(eV) | $\mu^a$ |
| CrC$_2$N$_4$ | Armchair | 637 | 3.08 | 8.11 | 13.35 | 1.38 | 2.99 | 918.55 |
| | Zigzag | 637 | 8.18 | 8.37 | 4.72 | 1.04 | 2.97 | 1235.31 |
| MoC$_2$N$_4$ | Armchair | 527 | 1.73 | 5.14 | 104.00 | 0.96 | 1.45 | 6299.23 |
| | Zigzag | 527 | 3.23 | 5.23 | 53.80 | 0.83 | 1.30 | 8508.35 |
| WC$_2$N$_4$ | Armchair | 546 | 1.37 | 6.65 | 68.92 | 0.77 | 1.89 | 5792.11 |
| | Zigzag | 546 | 5.66 | 5.67 | 22.75 | 0.67 | 1.68 | 8424.76 |

$^a$ Electron and hole mobilities at 298 K in unit of cm$^2$V$^{-1}$s$^{-1}$.

We next elaborately examine the phononic thermal transport MC$_2$N$_4$ (M= V, Cr, Mo, W) monolayers. We note that the for the case of VC$_2$N$_4$ nanosheet, because of its metallic electronic nature, phonon-electron interactions are ought to be considered for the complete understanding of thermal transport. Nonetheless, as a result of semiconducting nature of MC$_2$N$_4$ (M= Cr, Mo, W) monolayers, their predicted lattice thermal conductivities by the employed technique are expected to match closely with those measured experimentally [30]. The phonon dispersion relations of MC$_2$N$_4$ (M= V, Cr, Mo, W) monolayers are compared in Fig. 2. For the semiconducting monolayers we compared the MTP-based phonon dispersions with those acquired using the standard DFPT method, which show excellent agreements and confirm the outstanding accuracy of trained MTPs in the evaluation of interatomic force constants [27]. The temperature dependent lattice thermal conductivity of thermally and dynamically stable MC$_2$N$_4$ monolayers are plotted in Fig. 6a. The room temperature lattice thermal conductivity of CrC$_2$N$_4$, MoC$_2$N$_4$, WC$_2$N$_4$ and VC$_2$N$_4$ monolayers are predicted to be 350, 83, 79 and 37 W/mK, respectively, which highlight the substantial role of metal core atoms on the thermal transport. In accordance with structural, mechanical and electronic properties, MoC$_2$N$_4$ and WC$_2$N$_4$ nanosheets once again exhibit very close lattice thermal conductivities and conspicuously lower than that of the CrC$_2$N$_4$ counterpart. As a general rule, thermal conductivity decreases by temperature following a K~T$^{-\alpha}$ trend, where α is the temperature power factor. The temperature power factors for the thermal conductivity of CrC$_2$N$_4$, MoC$_2$N$_4$, WC$_2$N$_4$ and VC$_2$N$_4$ monolayers are found to be 1.35, 1.18, 1.22 and 1.05, respectively. Notably, the temperature power factor of CrC$_2$N$_4$ monolayer is remarkably close to that predicted for the single-layer graphene, 1.32 [43] and 1.34 [44]. The maximum α value for CrC$_2$N$_4$ nanosheet reveals stronger phonon-phonon interactions in this lattice which surge the thermal resistivity by increasing the temperature. Among the all studied nanosheets, the thermal transport in VC$_2$N$_4$ monolayer shows the lowest dependency to the temperature. In Fig. 6b, the cumulative phononic thermal conductivity as a function of frequency is plotted. These results show that the thermal conductivity increases steeply initially. It is moreover conspicuous that while the phonons with frequencies lower than 8 THz contribute to almost half of the total thermal conductivity, those with frequencies higher than 20 THz contribute negligibly. These observations reveal the dominance of acoustic phonons in the heat transport, however, yet the contribution of low-lying optical phonons is not negligible [45]. Our analysis of each phonon mode contribution reveal that acoustic modes yield 72, 80, 86 and 63 % of the total lattice thermal conductivity in CrC$_2$N$_4$, MoC$_2$N$_4$, WC$_2$N$_4$ and VC$_2$N$_4$ monolayers, respectively.

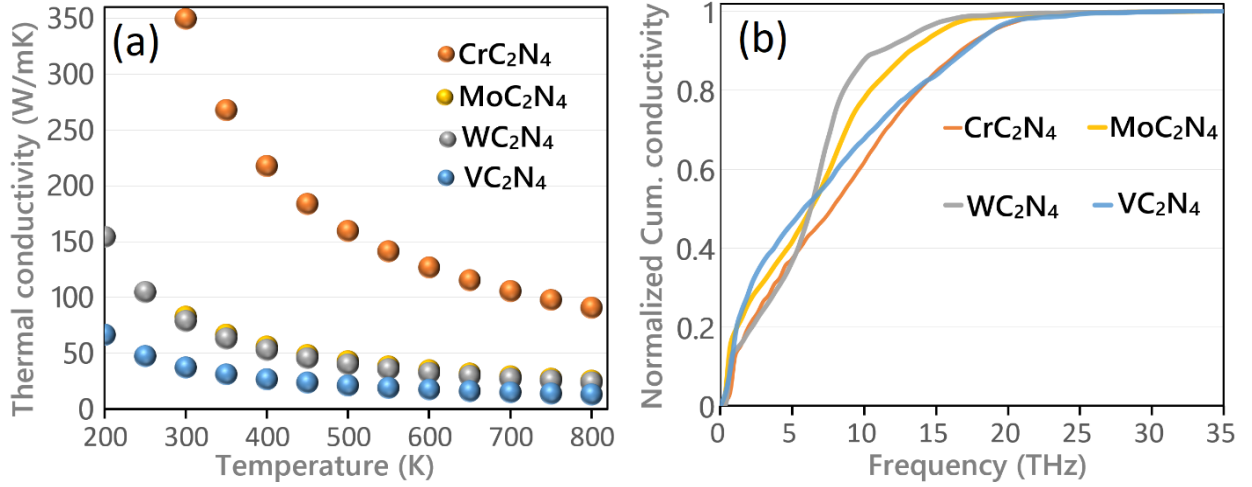

**Fig. 6**, (a) Temperature dependent lattice thermal conductivity and (b) normalized cumulative phononic thermal conductivity at 300 K as a function of the frequency for $CrC_2N_4$, $MoC_2N_4$, $WC_2N_4$ and $VC_2N_4$ monolayers.

The predicted trend for the lattice thermal conductivity of $MC_2N_4$ monolayers is not consistent with the classical theory, in which the thermal conductivity is expected to be higher for the systems with higher elastic modulus. To better understand the underlying mechanism in substantial role of metal core atoms on the resulting lattice thermal conductivity, we further examine the phononic properties. Due to the fact that acoustic phonons are the main heat carriers in these systems, their dispersions can provide useful information. From the basics, wider dispersion for a phonon mode reveals its faster group velocity which may result in a higher thermal conductivity. Moreover, the phonon bands crossing each other may surge the scattering and reduce the phononic thermal conductivity. In our investigation, we first compare the phonon group velocities for the studied monolayers, as shown in Fig. 7a and 7b. Presented results show distinctive decrease of the phonons' group velocities from $CrC_2N_4$ to $VC_2N_4$, $MoC_2N_4$ and $WC_2N_4$ monolayers. $CrC_2N_4$ monolayer exhibits the smallest lattice constant and less stretched C-N bonds among the considered systems, which results in wider dispersions for acoustic modes and consequently higher group velocities. Observed trend for phonons' group velocity is however not consistent with the estimated lowest thermal conductivity for $VC_2N_4$ monolayer. By considering the acoustic modes in the phonon dispersion relation of $VC_2N_4$ monolayer shown in Fig. 2, it is clear that they cross each other more frequently in comparison with the other studied systems, which promote the phonon scattering. In Fig. 7c and 7d we plot the phonons life time for the studied systems, which confirm the maximal scattering rate for the acoustic and optical modes in $VC_2N_4$ monolayer, as they show considerably lower life time. It is noticeable form Fig. 7d that acoustic low frequency modes in $WC_2N_4$ show generally higher life times that those in $MoC_2N_4$ counterpart, which partially compensate their lower phonon group velocity and finally result in very close lattice thermal conductivities in these 2D systems. The lattice thermal conductivity of $CrC_2N_4$ (350 W/mK) is in between of highly conductive metals of copper (385 W/mK) and gold (314 W/mK), but is still one order of magnitude lower than that of the single-layer graphene (~3600 W/mK). Interestingly, the predicted thermal conductivity for $CrC_2N_4$ is very close to that of the $CrSi_2N_4$ (332 W/mK [3]) and more than twice of $MoS_2$ monolayer [30]. The outstandingly high thermal conductivity of $CrC_2N_4$ is extremely appealing for the application in thermal management systems and minimizing the common overheating risks during the operation of electronic systems.

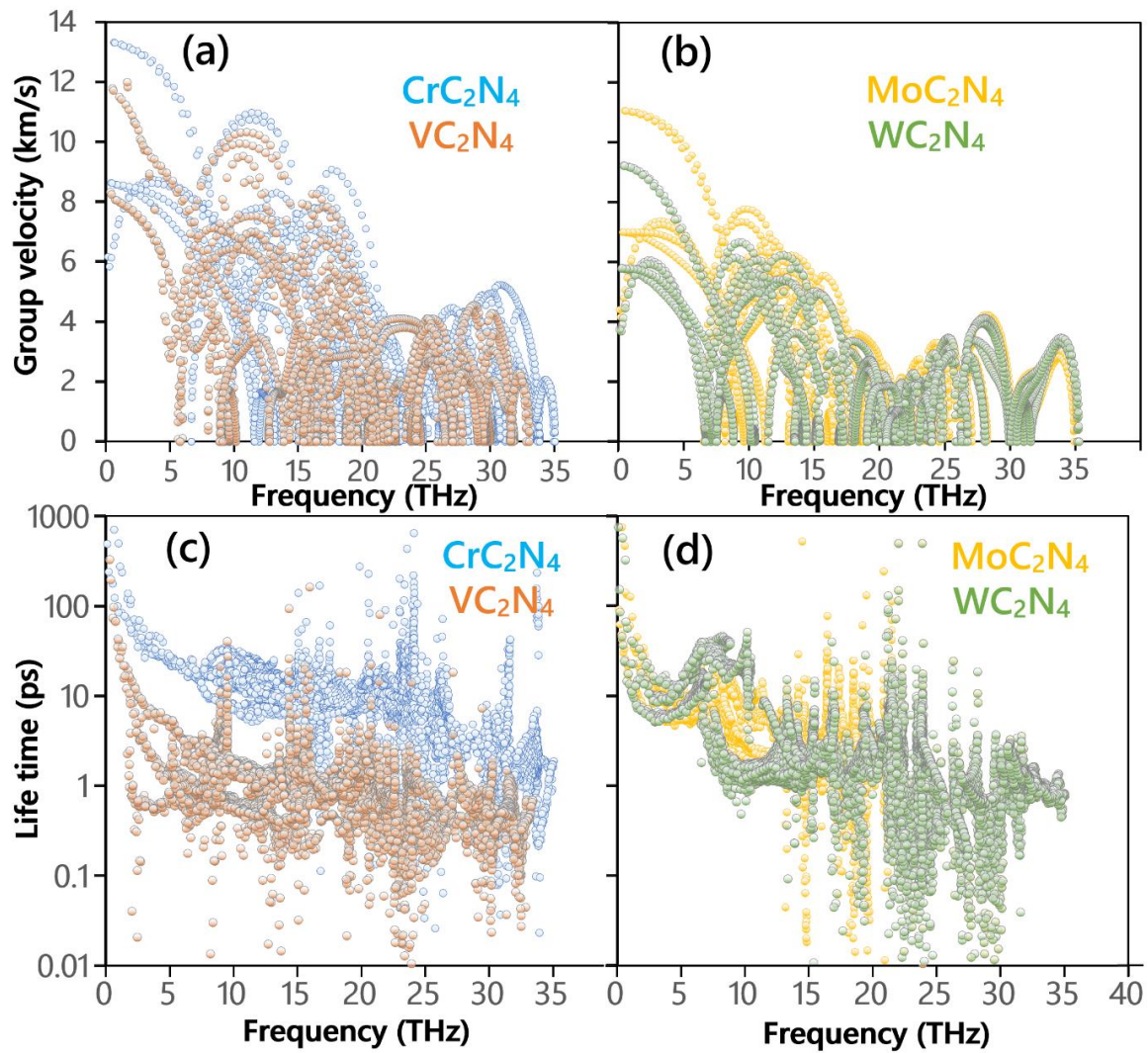

Fig. 7, (a, b) phonons' group velocities calculated with PHONOPY code (c, d) and phonons' life time obtained using the ShengBTE package of $CrC_2N_4$, $MoC_2N_4$, $WC_2N_4$ and $VC_2N_4$ monolayers.

Last but not least, we investigate the piezoelectric responses of semiconducting $CrC_2N_4$, $MoC_2N_4$ and $WC_2N_4$ monolayers using the CD model. The polarization estimated from CD model have contributions from both piezoelectric and flexoelectric effects, which is expressed as follows:

$$P_\alpha = d_{\alpha\beta}\epsilon_\beta + \mu_{\alpha\beta\delta}\frac{\partial \epsilon_\beta}{\partial r_\delta} \quad (3)$$

where $d_{\alpha\beta}$ is the piezoelectric coefficient and $\mu_{\alpha\beta\delta}$ corresponds to the flexoelectric coefficient. $\epsilon_{\beta\gamma}$ is the strain component and term $\frac{\partial \epsilon_\beta}{\partial r_\delta}$ represents the strain gradient in the atomic system. $\alpha, \beta$ and $\delta$ indicate the Cartesian coordinate directions. In order to estimate $d_{\alpha\beta}$, we have to separate the flexoelectric contribution from Eq. (3). For this, we calculate the different strain components associated with the given tensile deformation. Fig. 8a shows the various strain components $(\epsilon_\beta)$ for $CrC_2N_4$ monolayer at 10$^{th}$ load step with $K_t$=0.0005. Note that the strain values are measured using the OVITO package [46] strain analysis and then averaged over 30 equal width bins along $y$ axis of the atomic system. The non-zero strain components are $\epsilon_{yy}$ and $\epsilon_{zz}$. The prescribed loading induces $\epsilon_{yy}$ (from the first-order derivative of displacement field $u_y$ with $y$) as constant and make other strain components to zero, except the $\epsilon_{zz}$. The numerical values of $\epsilon_{yy}$ are nearly constant over all bins with a value of 0.005, which is 10 times the $K_t$. The $\epsilon_{zz}$ component arises due to the thermal relaxation and resulting changes in bond lengths and angles in the $z$ direction. Further it is noticed that $\epsilon_{zz}$ is constant during the tensile loading and the top and bottom portion symmetry of the monolayer is unaffected. This leads to a strong cancellation of $P_z$ dipole moments. As a result, the resultant polarization in $z$ direction, $P_y$ is zero (as shown in Fig. 8b). The other polarization component $P_x$ is also zero from Fig. 8b due to the non-existence of tensile loading induced strain $\epsilon_{xx}$. The bin averaged strain $\epsilon_{yy}$ is highly constant for the bins ranging from left to right of the given simulation box. Thus, the existence of gradient terms is very unlikely, which removes the terms related to strain gradients (flexoelectric parts) in the polarization Eq. 3. From the strain component to $y$ coordinate relationship, we confirm that there is only non-vanishing contribution from $\epsilon_{yy}$ to the polarization $P_y$. Hence, we establish the linear relationship for $P_y$ to $\epsilon_{yy}$ and estimate the $d_{yy}$ as piezoelectric coefficient.

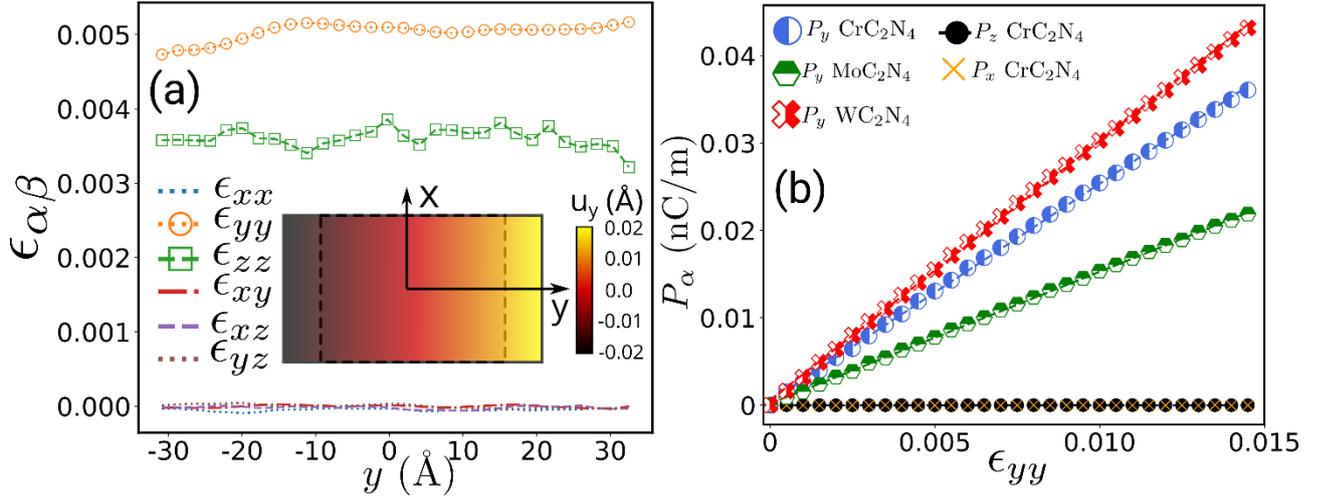

**Fig. 8,** (a) Distribution of strain components along $y$ direction for CrC$_2$N$_4$ monolayer for 20$^{th}$ load step with $K_t$=0.0005. The inset figure depict the tensile loading scheme with a defined color coding ranges from -0.02 to 0.02 Å for $u_y$. (b) The variation of polarization components ($P_\alpha$) with strain $\epsilon_{yy}$. Markers indicate the data from simulations and the lines indicate the linear curve fitting.

**Table 3,** Total polarizability estimated by DFT ($\alpha_{DFT}$) and the present CD model ($\alpha_{CAL}$). Atomic polarizability is R and $d_{yy}$ is the tensile piezoelectric coefficient.

| Lattice | $\alpha_{DFT}$ (Å$^3$) | $\alpha_{CAL}$ (Å$^3$) | $R_M$ (Å) | $R_A$ (Å) | $R_Z$ (Å) | $d_{yy}$ (nC/m) |
|---|---|---|---|---|---|---|
| CrC$_2$N$_4$ | 24.802 | 22.176 | 1.308 | 0.579 | 0.711 | 2.487 |
| MoC$_2$P$_4$ | 28.351 | 26.149 | 1.313 | 0.577 | 0.738 | 1.508 |
| WC$_2$N$_4$ | 34.758 | 34.585 | 1.405 | 0.549 | 0.729 | 2.966 |

Fig. 8b shows the linear variation between $P_y$ and $\epsilon_{yy}$ for CrC$_2$N$_4$, MoC$_2$N$_4$ and WC$_2$N$_4$ monolayers. The dipole moment depends on the effective atomic polarizability and the total electric field induced by the charges and dipoles. The first factor increases from, CrC$_2$N$_4$, MoC$_2$N$_4$ and WC$_2$N$_4$ as summarized in Table 3. However, $P_y$ is high for WC$_2$N$_4$, low for MoC$_2$N$_4$ and moderate for CrC$_2$N$_4$ at same strain state. The change of electric field $\Delta E_y$ between zero strain and at strain of 0.01 for CrC$_2$N$_4$ is 120.452 V/Å, MoC$_2$N$_4$ is 83.699 V/Å and WC$_2$N$_4$ is 140.595 V/Å, respectively. This shows that $P_y$ strictly follows the trend of the electric field. As the difference in electric field is more, the change of electronic configuration is more and helps to induce large dipole moments and thus polarization. Further this electric field has contributions from the charges $E_y^q$ and dipoles $E_y^p$. These electric fields demonstrate the effect of changes in the valance and bonding electrons in the generation of dipole moments via the $\pi - \sigma$ interactions and $\sigma - \sigma$ interactions. The changes in $\pi - \sigma$ interactions change the charge state of the atomic systems during the tensile deformation. The change of charge and the bonding distance to the neighbor atoms induces a local electric field $E_y^q$ [39]. For example, consider a unitcell in CrC$_2$N$_4$ monolayer, the change charge $\Delta q$ is 0.035$e$ between strain levels 0 to 0.01. For the unitcell selected at exactly similar locations for the MoC$_2$N$_4$ monolayer the $\Delta q$ becomes $0.022e$. The ratio of $\Delta q$ between CrC$_2$N$_4$ and the MoC$_2$N$_4$ exactly matches with the ratio of piezoelectric coefficients. The electron affinity of nitrogen is higher compared to chromium and carbon atoms, which means nitrogen atoms are reluctant to change their charge state. The difference in charge for atom labels B, D and F in Fig. 9a between strained and initial configuration is $0.002e$. The carbon atoms (labels C and E in Fig. 9a) bonded with the neighbor nitrogen atoms. The available electrons in carbon atoms are utilized in this bonding. Therefore, the given deformation slightly changed their charge state to $0.003e$, which is slightly higher than the nitrogen atoms due to the low electron affinity of carbon over nitrogen. The major change in the charge state appear for the chromium atom (label A in Fig. 9a). The low electron affinity of chromium helps to change the charge state to $0.008e$ during the tensile stretching. The total change in charge for CrC$_2$N$_4$ unitcell in Fig. 9a is $0.035e$. Also, the change in bond distances between carbon to nitrogen (B-C and E-F) are significantly lower than that of bond distance A-B and A-F. The change in bond distances in A-B is about 0.018 Å and A-F is about 0.005 Å. The difference in bond elongation between A-B and A-F bonds cancel the symmetry in generated electric fields and helps to produce the observed dipole moment. In the case of MoC$_2$N$_4$ monolayer, the bond elongations and change in charge states for nitrogen (labeled H, J and L in Fig. 9b) and carbon atoms (labeled as I and K in Fig. 9b) are similar to the CrC$_2$N$_4$ case. The only difference is that Mo has lower electron affinity compared to Cr and the bond elongations G-H and G-L differ only by 0.008 Å, which is 0.615 times lower than in CrC$_2$N$_4$ nanosheet. These low inducement of electric fields lowers the dipole moments and polarization for the Mo substitution. As a consequence of these effects, the piezoelectric coefficient for MoC$_2$N$_4$ is lower than that of the CrC$_2$N$_4$. In the case of WC$_2$N$_4$, the changes in bond elongations between the W and N atoms is about 0.013 Å, which is higher than MoC$_2$N$_4$ and slightly lower than CrC$_2$N$_4$. Though the polarization and piezoelectric coefficients for WC$_2$N$_4$ is higher than other two monolayers. This is mainly due to the inducement of electric field $E_y^p$. The difference in $E_y^p$ between initial and deformed state is about 23.532 V/Å, which is higher than that of CrC$_2$N$_4$ case (9.536 V/Å). The rise of $E_y^p$ is due to the interaction between the $\sigma - \sigma$ electrons which are associated with the changes in bond angles [47]. $\Delta\theta_2$ from Fig. 9c is about 0.299° and for $\Delta\theta_1$ from Fig. 9a is 0.266°. The increase in the $E_y^p$ helps to increase the polarization and the piezoelectric coefficient. Recently, SnOSe monolayer was predicted to show piezoelectric coefficient of 1.12 nC/m [48] with is notably the largest among the $MoS_2$ [39,49,50], XTeI [51] and Janus transition metal dichalcogenides [47,52]. Interestingly, the predicted piezocoefficients for $MC_2N_4$ (M= Cr, Mo, W) monolayers are distinctly higher than SnOSe. The predicted piezoelectric coefficient for $CrC_2N_4$ monolayer is about 10 orders higher than recently investigated Janus monolayers like group-III chalcogenides (0.325 nC/m) [13], group-IVB dichalcogendies (0.492 nC/m) [12], $Sb_2Se_2Te$ (out-of-plane 0.13 nC/m) [14], $Ga_2XY$ (0.231 nC/m) [11], $HfS_2Se$ (0.669 nC/m) [9] and SnSSe (0.2 nC/m)[10]. Nonetheless, piezocoefficients of these novel 2D systems are slightly lower than the those of $MSi_2N_4$ (M= Cr, Mo, W) [3] nanosheets.

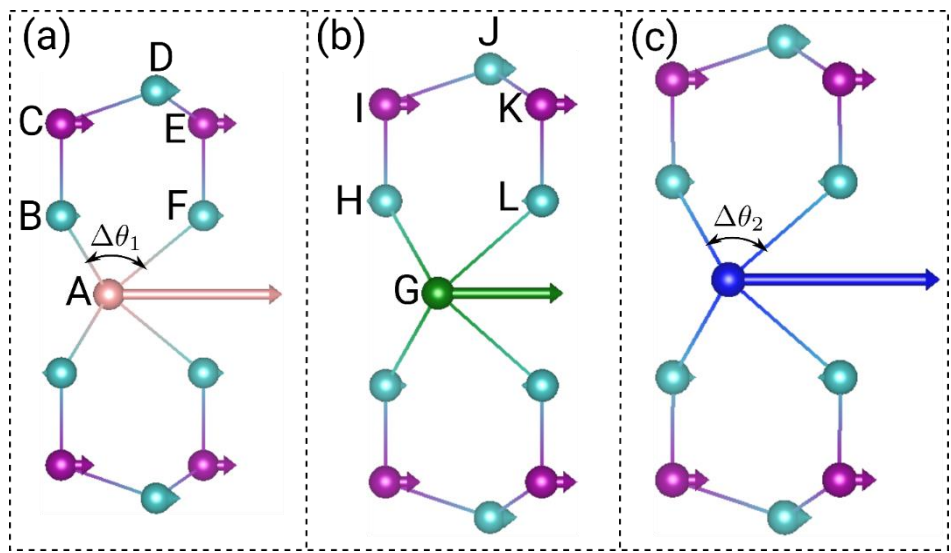

**Fig. 9**, Selected unitcell at a strain of 0.01 for (a) $CrC_2N_4$, (b) $MoC_2N_4$ and (c) $WC_2N_4$ monolayers. The arrows indicate the atomic dipole moments $p_y$.

4. Concluding remarks

Motivated by the latest experimental advance in the design and fabrication of single-layer $MoSi_2N_4$, herein we explore the stability and intrinsic properties of $MC_2N_4$ (M= Cr, Mo, W, V, Nb, Ta, Ti, Zr, Hf) monolayers. According to first-principles results, only $MC_2N_4$ (M= V, Cr, Mo, W) monolayers however show desirable thermal and dynamical stability. While $VC_2N_4$ lattice is shown to be a metallic system, $CrC_2N_4$, $MoC_2N_4$ and $WC_2N_4$ monolayers are interestingly found to be semiconductors with HSE06 band gaps of 2.32, 2.76 and 2.86 eV, respectively. With a direct band gap of 2.23 eV, a maximum carrier mobility of 1235 $cm^2V^{-1}s^{-1}$, and large absorption coefficients in the visible range of light, single-layer $CrC_2N_4$ offers excellent properties for the applications in nanoelectronics and optoelectronic. The elastic modulus of $CrC_2N_4$, $MoC_2N_4$, $WC_2N_4$ and $VC_2N_4$ monolayers are estimated to be remarkably high, 676, 543, 553 and 577 GPa, respectively. The maximum tensile strength of CrC$_2$N$_4$ nanosheet is predicted to be 54.8 GPa, which is only half of the that for graphene. The room temperature lattice thermal conductivity of CrC$_2$N$_4$, MoC$_2$N$_4$, WC$_2$N$_4$ and VC$_2$N$_4$ monolayers are predicted to be 350, 83, 79 and 37 W/mK, respectively, on the basis of iterative solution of the Boltzmann transport equation. CrC$_2$N$_4$ and WC$_2$N$_4$ lattices are noticeably found to exhibit high piezoelectric coefficients. Particularly CrC$_2$N$_4$ monolayer can show outstandingly high piezoelectric coefficient of 2.487 nC/m, more than twice of that in SnOSe, MoS$_2$ and XTeI monolayers. Our results reveal outstandingly high elastic modulus and tensile strength, lattice thermal conductivity, carrier mobility and piezoelectric coefficient in CrC$_2$N$_4$ nanosheet, highly appealing for the design of strong and stable electronic, thermal management and energy storage/conversion systems.

## Acknowledgment

B.M. and X.Z. appreciate the funding by the Deutsche Forschungsgemeinschaft (DFG, German Research Foundation) under Germany's Excellence Strategy within the Cluster of Excellence PhoenixD (EXC 2122, Project ID 390833453). F.S. thanks the Persian Gulf University Research Council for support of this study. B.J. and X.Z. gratefully acknowledge the sponsorship from the ERC Starting Grant COTOFLEXI (No. 802205). Authors also acknowledge the support of the cluster system team at the Leibniz Universität of Hannover. B. M and T. R. are greatly thankful to the VEGAS cluster at Bauhaus University of Weimar for providing the computational resources.

## Appendix A. Supplementary data

The following are the supplementary data to this article:

Supplementary Information

# Outstandingly high thermal conductivity, elastic modulus, carrier mobility and piezoelectricity in two-dimensional semiconducting CrC$_2$N$_4$: A first-principles study

Bohayra Mortazavi*,[a], Fazel Shojaei[b,#], Brahmanandam Javvaji[b,#], Timon Rabczuk[c] and Xiaoying Zhuang[a, c]

[a]*Chair of Computational Science and Simulation Technology, Institute of Photonics, Department of Mathematics and Physics, Leibniz Universität Hannover, Appelstraße 11,30167 Hannover, Germany.*
[b]*Department of Chemistry, Faculty of Sciences, Persian Gulf University, Boushehr 75168, Iran.*
[c]*College of Civil Engineering, Department of Geotechnical Engineering, Tongji University, 1239 Siping Road Shanghai, China.*

*E-mail: bohayra.mortazavi@gmail.com

VC2N4
 1.00000000000000
     2.5518395332987529    0.0000000000000000    0.0000000000000000
     1.2759197666493740    2.2099578621544600    0.0000000000000000
     0.0000000000000000    0.0000000000000000   22.0000000000000000
   V    N    C
     1    4    2
Direct
  0.3333333300000021  0.3333332999999996  0.5000003128615802
  0.6666666659999976  0.6666666666660035  0.5574175721153515
  0.6666666659999976  0.6666666666660035  0.4425818174745722
 -0.0000000000000000  0.0000000000000000  0.6453945613118636
  0.0000000000000000  0.0000000000000000  0.3546054928985894
  0.6666666659999976  0.6666666666660035  0.6220008709987448
  0.6666666659999976  0.6666666666660035  0.3779993723393127

CrC2N4
   1.00000000000000
     2.5130958697358361    0.0000000000000000    0.0000000000000000
     1.2565479348679180    2.1764048651960839    0.0000000000000000
     0.0000000000000000    0.0000000000000000   22.0000000000000000
   Cr   N    C
     1    4    2
Direct
  0.3333333300000021  0.3333332999999996  0.5000003128615802
  0.6666666659999976  0.6666666666660035  0.5553819725330271
  0.6666666659999976  0.6666666666660035  0.4446174170568966
  0.0000000000000000  0.0000000000000000  0.6437275105381488
  0.0000000000000000  0.0000000000000000  0.3562725436722900
  0.6666666659999976  0.6666666666660035  0.6205912952290049
  0.6666666659999976  0.6666666666660035  0.3794089481090523

MoC2N4
   1.00000000000000
     2.6223830538351329    0.0000000000000000    0.0000000000000000
     1.3111915269175629    2.2710503429402560    0.0000000000000000
     0.0000000000000000    0.0000000000000000   22.0000000000000000
   Mo   N    C
     1    4    2
Direct
  0.3333333300000021  0.3333332999999996  0.5000003128615802
  0.6666666659999976  0.6666666666660035  0.5619452983416551
  0.6666666659999976  0.6666666666660035  0.4380540912482616
  0.0000000000000000  0.0000000000000000  0.6502130310466825
  0.0000000000000000  0.0000000000000000  0.3497870231637492
  0.6666666659999976  0.6666666666660035  0.6267268805117325
  0.6666666659999976  0.6666666666660035  0.3732733628263460
WC2N4
   1.00000000000000
     2.6392440597375209    0.0000000000000000    0.0000000000000000

```
   1.3196220298687600    2.2856524023775999    0.0000000000000000
   0.0000000000000000    0.0000000000000000   22.0000000000000000
  W    N    C
   1    4    2
Direct
 0.3333333300000021  0.3333332999999996  0.5000003128615802
 0.6666666659999976  0.6666666666660035  0.5617929719118555
 0.6666666659999976  0.6666666666660035  0.4382064176780539
-0.0000000000000000  0.0000000000000000  0.6502327087123897
-0.0000000000000000  0.0000000000000000  0.3497673454980492
 0.6666666659999976  0.6666666666660035  0.6267679574664891
 0.6666666659999976  0.6666666666660035  0.3732322858715965
```

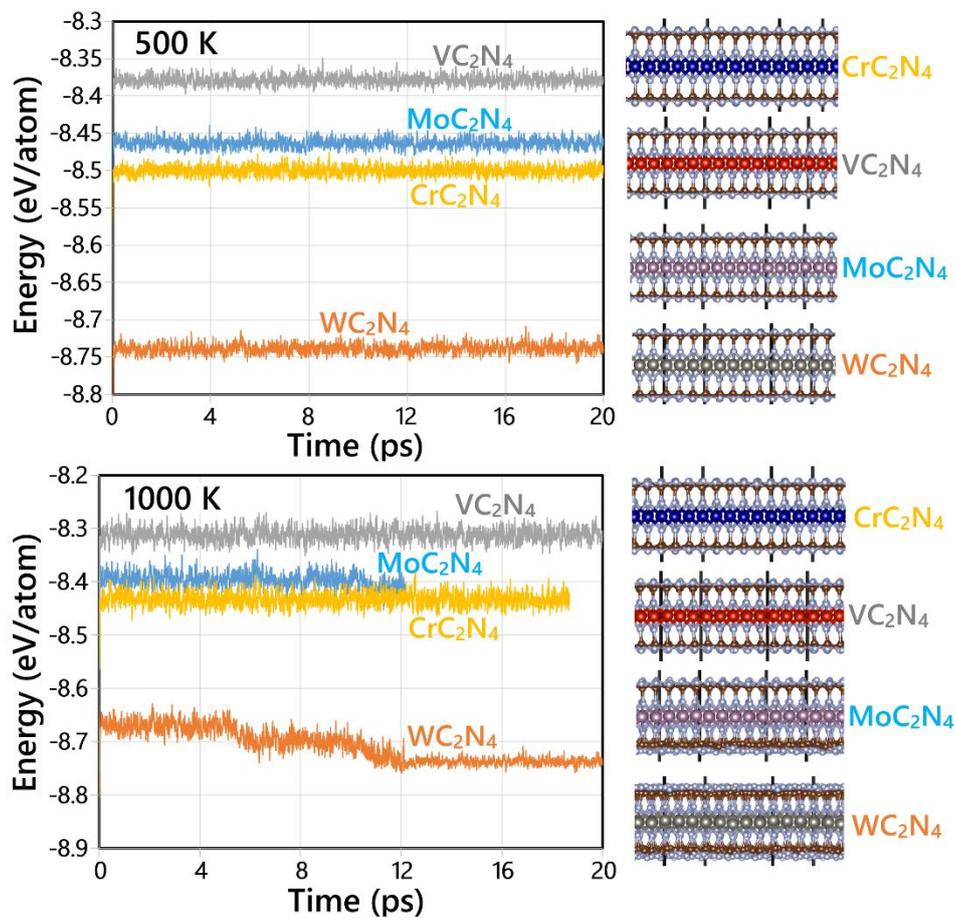

**Fig. S1**, Evolution of potential energy during the AIMD simulations for 2H-MC$_2$N$_4$ (M= Cr, Mo, W) monolayers at 500 and 1000 K. The atomic lattices at the end of simulations are depicted in the right side.

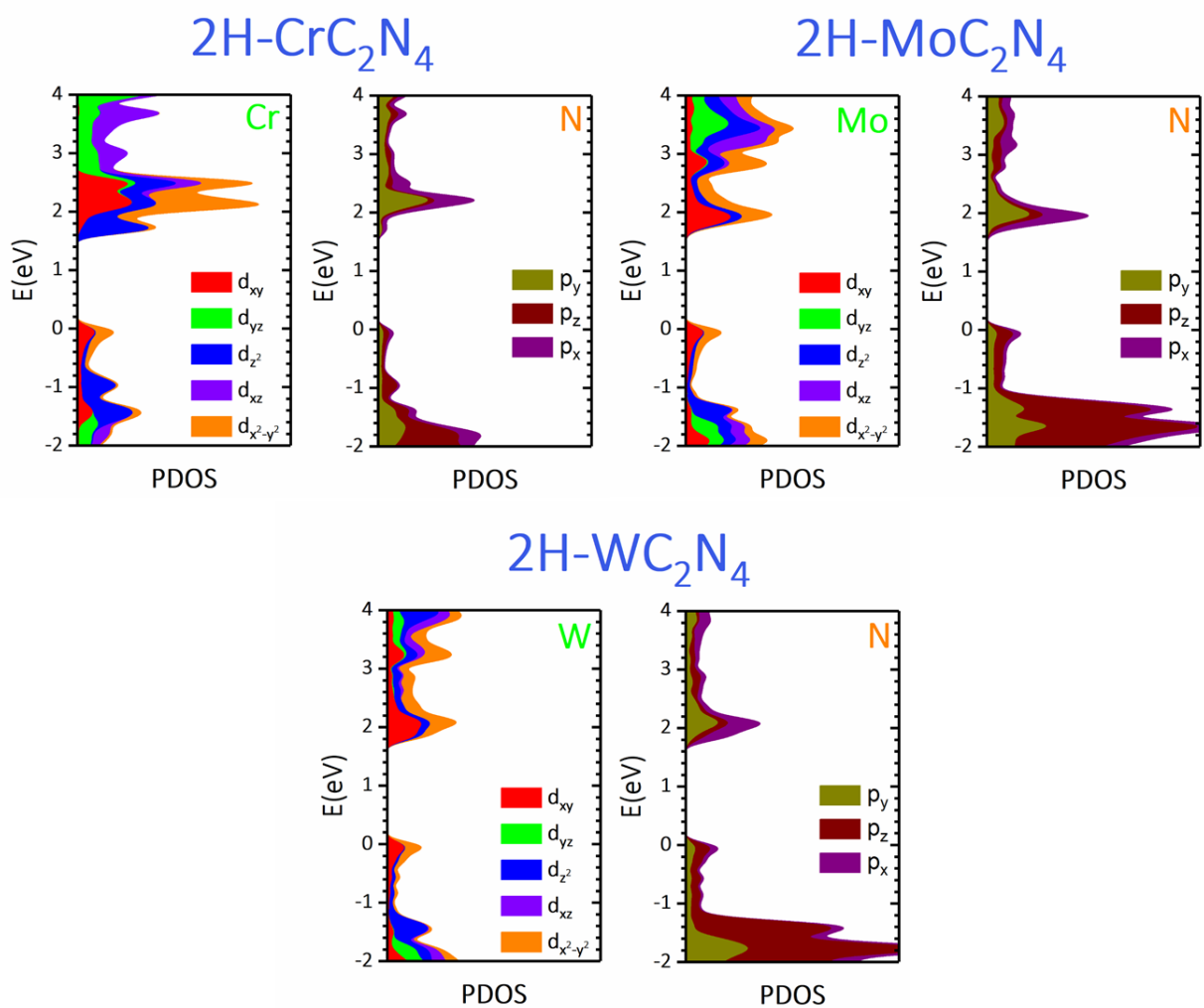

Fig. S2, Orbital projected density of states for M and N atoms of 2H-MC$_2$N$_4$ (M= Cr, Mo, W) monolayers. The Fermi level is set to zero.

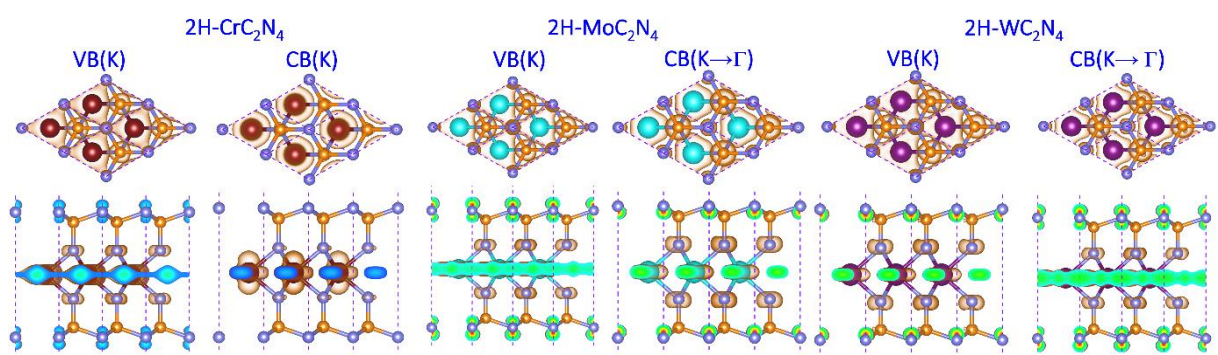

Fig. S3, Charge density distributions of 2H-MC$_2$N$_4$ (M= Cr, Mo, W) monolayers at their VBMs and CBMs.